\PassOptionsToPackage{unicode}{hyperref}
\PassOptionsToPackage{hyphens}{url}
\documentclass[
]{article}
\usepackage{lmodern}
\usepackage{amsmath}
\usepackage{ifxetex,ifluatex}
\ifnum 0\ifxetex 1\fi\ifluatex 1\fi=0 
  \usepackage[T1]{fontenc}
  \usepackage[utf8]{inputenc}
  \usepackage{textcomp} 
  \usepackage{amssymb}
\else 
  \usepackage{unicode-math}
  \defaultfontfeatures{Scale=MatchLowercase}
  \defaultfontfeatures[\rmfamily]{Ligatures=TeX,Scale=1}
\fi
\IfFileExists{upquote.sty}{\usepackage{upquote}}{}
\IfFileExists{microtype.sty}{
  \usepackage[]{microtype}
  \UseMicrotypeSet[protrusion]{basicmath} 
}{}
\makeatletter
\@ifundefined{KOMAClassName}{
  \IfFileExists{parskip.sty}{%
    \usepackage{parskip}
  }{
    \setlength{\parindent}{0pt}
    \setlength{\parskip}{6pt plus 2pt minus 1pt}}
}{
  \KOMAoptions{parskip=half}}
\makeatother
\usepackage{xcolor}
\IfFileExists{xurl.sty}{\usepackage{xurl}}{} 
\IfFileExists{bookmark.sty}{\usepackage{bookmark}}{\usepackage{hyperref}}
\hypersetup{
  hidelinks,
  pdfcreator={LaTeX via pandoc}}
\ifluatex
  \usepackage{selnolig}  
\fi

\date{}

\usepackage{graphicx}
\usepackage{hyperref}
\usepackage{pdflscape} 
\usepackage{multirow}
\usepackage{setspace}
\usepackage{lineno}
\usepackage{authblk}
\usepackage{soul}
\usepackage{pdfpages}
\usepackage{fancyhdr}
\usepackage{booktabs}
\usepackage{longtable}

\title{Spatial shifts in productivity of the coastal ocean over the past two
	decades induced by migration of the Pacific Anticyclone and Bakun effect in the Humboldt Upwelling Ecosystem\footnote{© 2020. This manuscript version is made available under the CC-BY-NC-ND 4.0 license \href{http://creativecommons.org/licenses/by-nc-nd/4.0/}{http://creativecommons.org/licenses/by-nc-nd/4.0/}}}

\author[1]{Nicolas Weidberg\footnote{Current address: University of South Carolina, Department of Biological Sciences, Columbia, South Carolina, USA.}}

\author[2]{Andres Ospina-Alvarez\footnote{Corresponding author: Andrés Ospina-Alvarez, email: aospina.co@me.com; address: Spanish Scientific Research Council, Mediterranean Institute for Advanced Studies (IMEDEA-CSIC/UIB), C/ Miquel Marques 21, CP 07190 Esporles, Balearic Islands, Spain.}}

\author[3]{Jessica Bonicelli}

\author[4]{Mario Barahona}

\author[1]{Christopher M. Aiken}

\author[4]{Bernardo R. Broitman}

\author[1]{Sergio A. Navarrete}

\affil[1]{Estación Costera de Investigaciones Marinas, Las Cruces, and
	Center for Applied Ecology and Sustainability (CAPES), 6513677,
	Casilla 193, Correo 22, Santiago, Chile.}
\affil[2]{Mediterranean Institute for Advanced Studies (IMEDEA-CSIC/UIB), C/ Miquel Marques 21, CP 07190 Esporles, Balearic Islands, Spain.}
\affil[3]{Instituto de Fomento Pesquero (IFOP), Almte. M. Blanco Encalada 839, Casilla 8-V, Valparaiso, Chile.}
\affil[4]{Departamento de Ciencias, Facultad de Artes Liberales Liberales $ \& $ Bioengineering Innovation Center, Facultad de Ingeniería y Ciencias, Universidad Adolfo Ibáñez, Viña del Mar, Chile.}

\begin{document}
	\maketitle
	
	\textbf{Running page head:} Global change and ocean primary productivity

\begin{abstract}

Intensification and poleward expansion of upwelling-favourable winds have been predicted as a response to anthropogenic global climate change and have recently been documented in most Eastern Boundary Upwelling Ecosystems of the world. To identify how these processes are impacting nearshore oceanographic habitats and, especially, long-term trends of primary productivity in the Humboldt Upwelling Ecosystem (HUE), we analysed time series of sea level pressure, wind stress, sea surface and atmospheric surface temperatures, and Chlorophyll-a, as a proxy for primary productivity, along 26º--36° S. Major artisanal and industrial
fisheries are supported by phytoplankton productivity in this region
and, therefore, identification of long-term trends and their spatial
variability is critical for our ability to adapt to and to mitigate the
effects of global climate change. We show that climate-induced trends in
primary productivity are highly heterogeneous across the region. On the
one hand, the well-documented poleward migration of the South Pacific
Anticyclone (SPA) has led to decreased spring upwelling winds in the
region between ca. 30° and 34° S, and to their intensification to the
south. Decreased winds have produced slight increases in sea surface
temperature and a pronounced and meridionally extensive decrease in
surface Chlorophyll-a in this region of central Chile. To the north of
30° S, significant increases in upwelling winds, decreased SST, and
enhanced chlorophyll-a concentration are observed in the nearshore. We show that this increased in upwelling-driven coastal productivity is
probably produced by the increased land-sea pressure gradients (Bakun's
effect) that have occurred over the past two decades north of 30° S.
Thus, climate drivers along the HUE are inducing contrasting trends in
oceanographic conditions and primary productivity, which can have
far-reaching consequences for coastal pelagic and benthic ecosystems and
lead to geographic displacements of the major fisheries.

	\end{abstract}


\section{Introduction}

Documenting long-term trends in Eastern Boundary Upwelling Systems (EBUSs) is of great importance not only because a large fraction of the human population lives near these shores, but also because these are the most productive marine regions in terms of phytoplankton biomass and fish stocks (Cushing 1971; Cury et al., 1998; Chavez and Messie, 2009). Indeed, equator-ward winds along these shores induce the upwelling of subsurface cold water masses rich in nutrients, which fuel phytoplankton growth and, through the trophic web, provide the necessary energy to sustain some of the largest industrial and artisanal fisheries of the world (Mann and Lazier, 2006; Chenillat et al., 2013; Chavez and Messie, 2009; Salas et al., 2011). Thus, variability of climatic drivers on surface water productivity within EBUS regions impinges directly on our ability to adapt to climate change and on the sustainability of major industrial and local fisheries, with the large direct and indirect socio-economic impacts they generate (Pauli and Christensen, 1995). Consequently, several recent studies have examined long-term trends in wind forcing and physical oceanographic conditions in most EBUSs and, naturally, our understanding of long-term climate effects in these regions of the worlds' oceans has improved considerably. The effects that these long-term hydrographic changes are having on productivity of the coastal oceans are now a matter of intensive research. However, to understand these temporal trends in primary productivity, studies simultaneously integrating large scale climatic forcing, wind dynamics and sea surface temperature variability are still lacking for the most important EBUSs.

On a global scale, the poleward expansion of the atmospheric Hadley circulation cells, due to the weakening of the thermal differences
between the poles and the equator, is known to have already caused the poleward displacement of both the subtropical anticyclones and the
sub-polar westerlies belts (Previdi and Liepert, 2007; Nguyen et al., 2013). The result has been a poleward expansion and intensification of
upwelling favourable winds along most EBUS over the past two decades
(McGregor et al., 2007; Lima and Wethey, 2012; Sydeman et al., 2014).
The trend is particularly evident in the Southern Pacific, along the
Humboldt Upwelling Ecosystem (HUE), where the poleward movement of the
Southwestern Pacific Anticyclone (SPA) has displaced the westerlies belt
closer to Antarctica (Fan et al, 2014; Ancapichún and Garcés-Vargas
2015). Recent studies have shown that the seasonal latitudinal migration
of the SPA has moved poleward to sit around 36°S during the critical
austral spring-summer months that fuel phytoplankton productivity in the
coastal section of the south eastern Pacific ocean (Schneider et al.,
2017; Aguirre et al., 2018). The displacement has been captured by one
of the main modes of climatic variability in the region, the Pacific
Annular Mode (SAM), which has shown a clear positive trend in the last
decades, thus pointing to low atmospheric pressures over Antarctica
compared to those at subtropical latitudes (Marshall, 2003; Wang and
Cai, 2013). It has also driven a shift in atmospheric conditions along a
broad region of the central coast of Chile, between ca. 30-35° S, where
spring winds have persistently decreased over the past decades (Aguirre
et al., 2018), and a poleward region where winds have intensified,
rainfall has decreased and surface water cooling has been observed
throughout the mixed layer, especially since 2007 (Schneider et al.,
2017; Jacob et al., 2018; Narváez et al., 2019). The latitudinal
position of the SPA and its regular seasonal migration clearly modulate
long-term trends in upwelling phenology and atmospheric climatic
conditions (Ancapichún and Garcés-Vargas, 2015; Weller 2015, Jacob et
al., 2018), and impose geographic discontinuities in upwelling regimes,
which in the HUE occur around 30° S (Strub et al., 1998; Hormazabal et
al, 2001; Navarrete et al., 2005; Tapia et al., 2014). Such
discontinuities in upwelling regimes are known to have farreaching
consequences on pelagic and benthic ecosystems, including population
dynamics, larval recruitment, adult abundances, the role of species
interactions, and genetic and functional structure of benthic
communities (Navarrete et al., 2005; Wieters et al., 2009; Tapia et al.,
2014; Haye et al., 2014). How climate forcing and the progressive
poleward displacement of the SPA will in the long-term affect the
biogeographic discontinuity at 30°S remains unclear.

Besides the intensification and poleward migration of subtropical
anticyclones, in 1990 Bakun (Bakun 1990) predicted that the greenhouse
effect would warm the oceans slower than adjacent land masses, thus the
land-sea pressure differential will become reinforced, strengthening
landward breezes which, due to Coriolis, would increase equatorward
winds and coastal upwelling at all EBUSs. The phenomenon is known as the
Bakun effect (Bakun 1990; Bakun et al., 2010), and empirical evidence
and model results supporting such response to anthropogenic warming have
been controversial (Demarcq 2009; Pardo et al., 2011). Recent
meta-analyses of observations and models show that significant
intensification of winds, attributable to Bakun effect, has occurred in
all EBUSs but the Canary Current system, where winds have weakened
(Sydeman et al., 2014). This result is consistent with several in-situ
observations (García-Reyes and Largier, 2010; Weller 2015) and model
predictions for the XXI century (Wang et al. 2015). But other studies,
using different databases, have found no evidence for a general
intensification of upwelling along EBUSs (Varela et al., 2015).
Contradictory results obtained by different studies are due, at least in
part, to the fact that long-term responses within EBUSs are not
homogeneous, but trends may change across latitudes (Varela et al.,
2015). Thus, the regional averages examined to date may provide an
incomplete picture of long-term trends in the highly dynamic and
spatially variable coastal ocean that characterize all EBUS. 

To what extent the positive or negative trends in equatorward winds, the
driving force of upwelling, have led to changes in primary productivity
of coastal waters is now being intensively investigated (Bakun et al.,
2015; Gomez-Letona et al., 2017). The effect of altered upwelling winds
on surface primary productivity can have contrasting effects over
coastal and offshore domains and analyses must therefore take this into
account. For instance, stronger coastal upwelling generates turbulence
and mixing alongshore which, together with enhanced seaward transport by
the Ekman drift, may lead to higher offshore phytoplankton
concentrations at the expense of a decrease in nearshore primary
productivity (Lackhar and Gruber, 2012; Anabalon et al., 2016). In
contrast, a scenario with higher phytoplankton biomass nearshore induced
by enhanced upwelling-driven nutrient inputs is also possible (Bakun et
al., 2010). Moreover, the influence of upwelling intensification or
weakening on phytoplankton biomass may depend on the ecophysiological
requirements of different phytoplankton groups (Smayda, 2000) or changes
in other nutrient sources, such as riverine inputs (Masotti et al.,
2018; Jacob et al., 2018).

In this study, we aim to quantify and disentangle the role of SPA
dynamics and Bakun effect on upwelling spatiotemporal fluctuations at
the Humboldt Upwelling System. To this end, time series of sea-land
thermal differentials, sea level atmospheric pressure, meridional wind
stress, satellite sea surface temperature (SST) and chlorophyll-a
concentration (Chl-a) with the highest possible spatiotemporal
resolution for the region, together with \emph{in situ} measurements of
nearshore Chl-a, were inspected along central Chile between 26º--36° S
in search of interannual and seasonal trends. With this approach we were
able to evaluate the strength of such trends, their spatial distribution
across the region and their correspondence with the displacement of the
SPA and Bakun effect. The direction and spatial structure of the
temporal trends measured in all these variables will allow us further
comparisons with other models and observations developed both within the
Humboldt Upwelling System and in other EBUSs. 

\section{Material and Methods}

\subsection*{Physical Conditions: Winds, SPA migration and Bakum'{~ }effect}
Processed, science-quality satellite data for SST and Chl-a were retrieved from the Moderate Resolution Imaging Spectroradiometer (MODIS) onboard the Aqua spacecraft\footnote{{https://coastwatch.pfeg.noaa.gov/erddap/griddap/erdMH1sstd8day.html}; {https://coastwatch.pfeg.noaa.gov/erddap/griddap/erdMH1chla8day.html}}. In particular, 8day composites with 4 km of pixel resolution were
retrieved at different distances from the shore (20, 40, 60, 80, 100, 200, 300, 400 and 500 km) between 26º--36° S off the Chilean coast from
May 2003 to December 2015. To reduce the spatial variability in the
datasets and the gaps along the time series, latitudinal moving averages
from North to South every 4 km using pixel aggregations (superpixels)
with a length of 12 km alongshore  ·  20 km cross shore (3  ·  5 pixels)
were calculated and used in further analyses.

The ERA-Interim weather reanalysis model from the European Center for
Medium-Range Weather Forecast ({http://apps.ecmwf.int/datasets/}) was
used to obtain meridional wind velocities at a height of 10 m
(\emph{V}). Input data for wind data prediction and reanalysis is
provided by Meteosat satellites (Dee et al., 2011). Model estimates were
downloaded as monthly means for pixels of 0.75° from 26.25º S to 36°S
and from 73º W to 69.75° W. Then, meridional wind stress, the component
which drives upwelling along the Chilean coast, was calculated asas it
has been{~ }previously used to characterize Ekman transport dynamics in
the region (Aguirre et al. 2014). It was computed as follows:

\begin{equation}\label{Equation 1}
	W_{stress}=\sigma_{air} C D V^2
\end{equation}

where W\textsubscript{stress} stands for wind stress,
$ \sigma $ \textsubscript{air} is air density (1.225 kg m\textsuperscript{-3}), CD
is the drag coefficient (0.0014) and \emph{V} is meridional wind speed.
Positive and negative values represent northwards and southwards wind
forcing, respectively.

In order to infer SPA position over time, from the ERA Interim model
monthly sea level atmospheric pressures at pixels with a size of 0.75°
were retrieved from 2003 to 2015 in between 76º--96° W and 25º--40° S.
Then, a third-order polynomial fit was applied to sea level pressures as
a function of latitude following the method of Minetti et al. (2009).
The SPA position was calculated for the entire time series as the
latitude where the maximum of the polynomial curve (first derivative
zero) is reached.

To evaluate Bakun effect across the study region over the timespan
encompassed by our study, monthly ERA Interim modelled surface
atmosphere (`skin') temperatures at pixels with a size of 0.75° were
retrieved from 2003 to 2015 from the first inland pixel in between
26º--36° S. Then, the monthly sea-land thermal differences (in °C) was
obtained by subtracting monthly average SST values at the 20 km
superpixel and over 0.75° latitudinal bands, from the monthly land skin
temperatures.

\subsection*{Chlorophyll-a in surface waters: in situ and satellite data}

Surface Chl-a data were retrieved from the Moderate Resolution Imaging
Spectroradiometer (MODIS) at the Aqua spacecraft using the same
procedure described above for SST. 

To validate satellite data and examine in-situ time trends, we used the
time series of daily, Chlorophyll-a extracted at the Estación Costera de
Investigaciones Marinas (ECIM) located at 33.49° S (Fig. 1). Three 250
ml water samples were collected daily at depths between 20--40 cm from
2003 to 2012 and filtered onto 0.2 um filters. Then these were
prefiltered with a $ 300 \mu m $ mesh and $ 100 ml $ were filtered through a GF/F Whatman glass fibre filter (Wieters et al., 2003). The Chl -a was
extracted by adding 5 ml of 90\% acetone during 24 h at -20 °C to the
GF/f filters (Yentsch and Menzel, 1963). Concentrations were then
measured with a 10-AU Turner Designs fluorometer and used to obtain
daily averages for a time series. 

\begin{figure}
	\centering
	\includegraphics[width=1\linewidth]{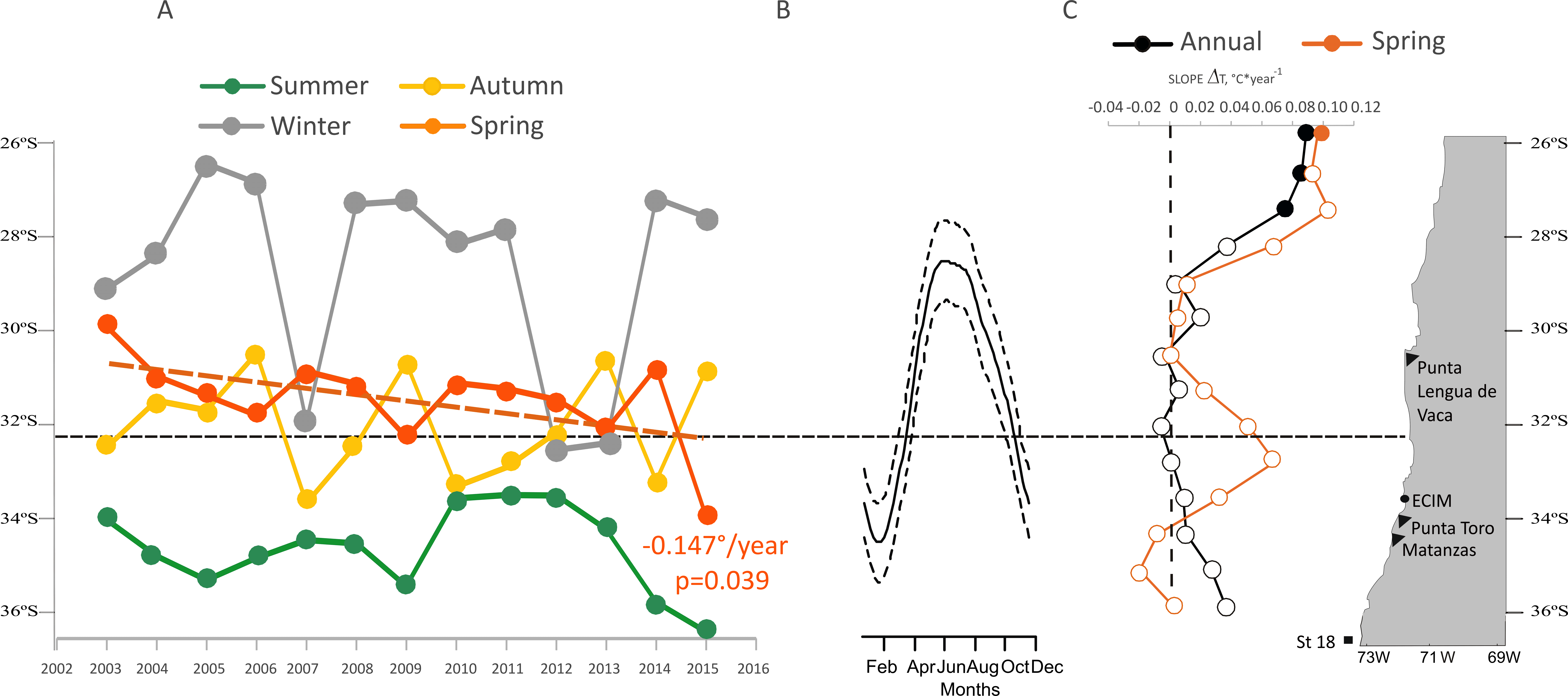}
	\caption{Map of the study area. A) Time series of seasonal latitudinal positions of the SPA showing the significant negative trend in spring (orange discontinuous line). B) SPA position inferred from the seasonal component of the GAMM applied to monthly sea level pressure data, with dashed lines representing 95\% confidence interval. C) Latitudinal trends of long-term slopes of sea-land thermal difference for the whole year and in spring. Filled circles show significant trends. The straight dashed line across the three panels represents the mean position of the SPA.}
	\label{fig:fig1}
\end{figure}

\subsection*{Data Analyses}

Generalized Additive Mixed Models (GAMMs, Chen, 2000; Wood, 2006) were
used to decompose time series of wind stress estimates, SPA positions,
sea-land thermal differences, SST, Chl-a, and \emph{in situ} Chl-a at
ECIM, into a seasonal and a long-term trend component. Like standard
Generalized Additive Models, GAMM fits a penalized cubic spline
smoothing functions through backfitting procedures, but it allows to
account for temporal autocorrelation and lag effects in the error
structure through the inclusion of an autoregressive term, which is
considered a random effect in the GAMM. The model is robust to data gaps
and irregular time intervals between measurements along the time
series

(Simpson, 2018). These analyses were implemented with the ``mgcv''
package (Wood, 2006) for R statistical computing version 3.4.0 (R Core
Team 2019) and have been used before to{~ }analyse time series of
hydrographical and biological variables (Harding et al., 2016). A linear
fit was applied to the long-term trend component to obtain a time
integrated slope or rate of change for wind stress, SST and Chl-a. In
order to infer spatial patterns in temporal trends these slopes were
calculated for each smoothed superpixel of 12 · 20 km in the case of
MODIS data and for each 0.75° pixel in the case of ERA Interim wind
stress estimates and represented in maps for the whole region. 

To examine the coherence of the long-term trends among seasons,
especially during the period of maximum primary productivity in spring,
linear trends for each season were performed separately for each wind
stress pixel and MODIS superpixel. Seasonal means for each variable were
obtained using the following periods: spring (September to November),
summer (December to February), autumn (March to May) and winter (June to
August). Seasonal slopes of these linear trends were also represented in
maps for the whole region. 

Since previous studies have reported that poleward displacement of SPA
has accelerated since 2007 (Schneider et al., 2017; Aguirre et al.,
2018), mean seasonal sea level pressure maps for the periods 2003--2007
and 2008--2015 were compared in search of contrasting spatial patterns
in the location of the SPA. We also calculated the difference in
meridional wind stress, SST and Chl-a between the two periods for those
coastal areas showing the most contrasting long term trends in Chl-a
within the study domain. Chl-a and SST data were averaged for the first
100 km offshore at these coastal areas and the respective 0.75° pixels
were selected to calculate the differentials of wind stress between the
periods before and after 2007. These differentials were compared with
those observed in previous studies further south at 36.5° S (Schneider
et al., 2017; Aguirre et al., 2018).

\section{Results}

Time series of monthly SPA latitudinal position revealed{~ }strong
seasonality, with the core of the anticyclone located at 29° S in
austral winter (July and August) and at 35° S at the end of austral
summer (Fig. 1B). The GAMM analyses revealed a slight southward trend
along the entire time series, and within each season, linear regressions
showed sharp and significant poleward displacement in spring (Fig. 1A),
as the SPA shifted from 31° S in 2003 to 35° S in 2015. Given this sharp
spring trend, further analyses on sea-land thermal contrasts and
satellite datasets focused on this season. Detailed analyses of sea
level pressure in the region during the austral upwelling spring
transition for the period before (2003-2007) and the period after 2007 provide a visual confirmation of these changes (Fig 2A-D). After 2007 sea level pressures in spring at the core of the SPA have increased, and the position is more skewed towards the south (Fig. 2D).

\begin{figure}
	\centering
	\includegraphics[width=1\linewidth]{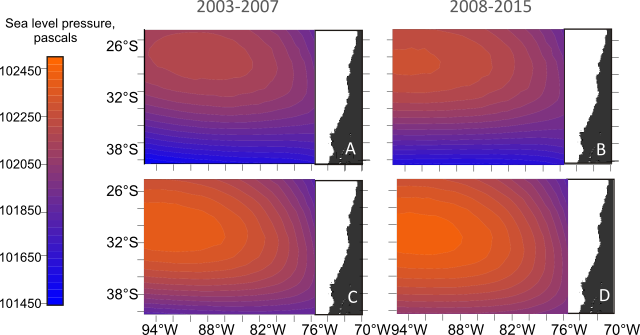}
	\caption{Contour maps of sea level pressure in winter (A, B) and spring (C, D) before (A, C) and after 2007 (B, D).}
	\label{fig:figura-2}
\end{figure}

We observed highly significant positive long-term trends in sea-land
thermal differences (i.e. Bakun effect), yet the temporal pattern showed
a clear latitudinal discontinuity. North of about 30° S, a significant
increase took place in both annual and spring sea-land pressure
gradient, with a slope of about 0.08° C year\textsuperscript{-1} over
the timespan considered. South of 30° S, annual pressure differential
showed no trend and the slope remained close to 0, while spring pressure
differential showed a non-significant positive trend at some latitudinal
bands (Fig.1C). 

On average, across the entire study region, long-term trends of
meridional wind stress were mostly positive for the entire year at
around 0.001 N m\textsuperscript{-2} year\textsuperscript{-1}, with
slightly steeper trends north from 28° S (Fig 3A) and nearly null or
even slightly negative trends (-0.0005 N m\textsuperscript{-2}
year\textsuperscript{-1}) close to shore between 30º--34° S. During the
spring upwelling months, two contrasting regions in terms of trends in
wind stress were clearly defined. Firstly, a region or domain north of
30°S, where marked increases in meridional wind stress took place at a
rate of more than 0.002 N m\textsuperscript{-2} year\textsuperscript{-1}
at 27° S. Secondly, a region or domain south of 30° S to 35° S, where
meridional winds decreased at rates around 0.0015 N
m\textsuperscript{-2} year\textsuperscript{-1} (Fig. 3E).
Intensification of upwelling winds in the northern domain also took
place during summer, while in autumn the increasing trend was only
evident offshore (\textless100 km off the coast) in between 31º S to 34°
S (S1).

\begin{figure}
	\centering
	\includegraphics[width=1\linewidth]{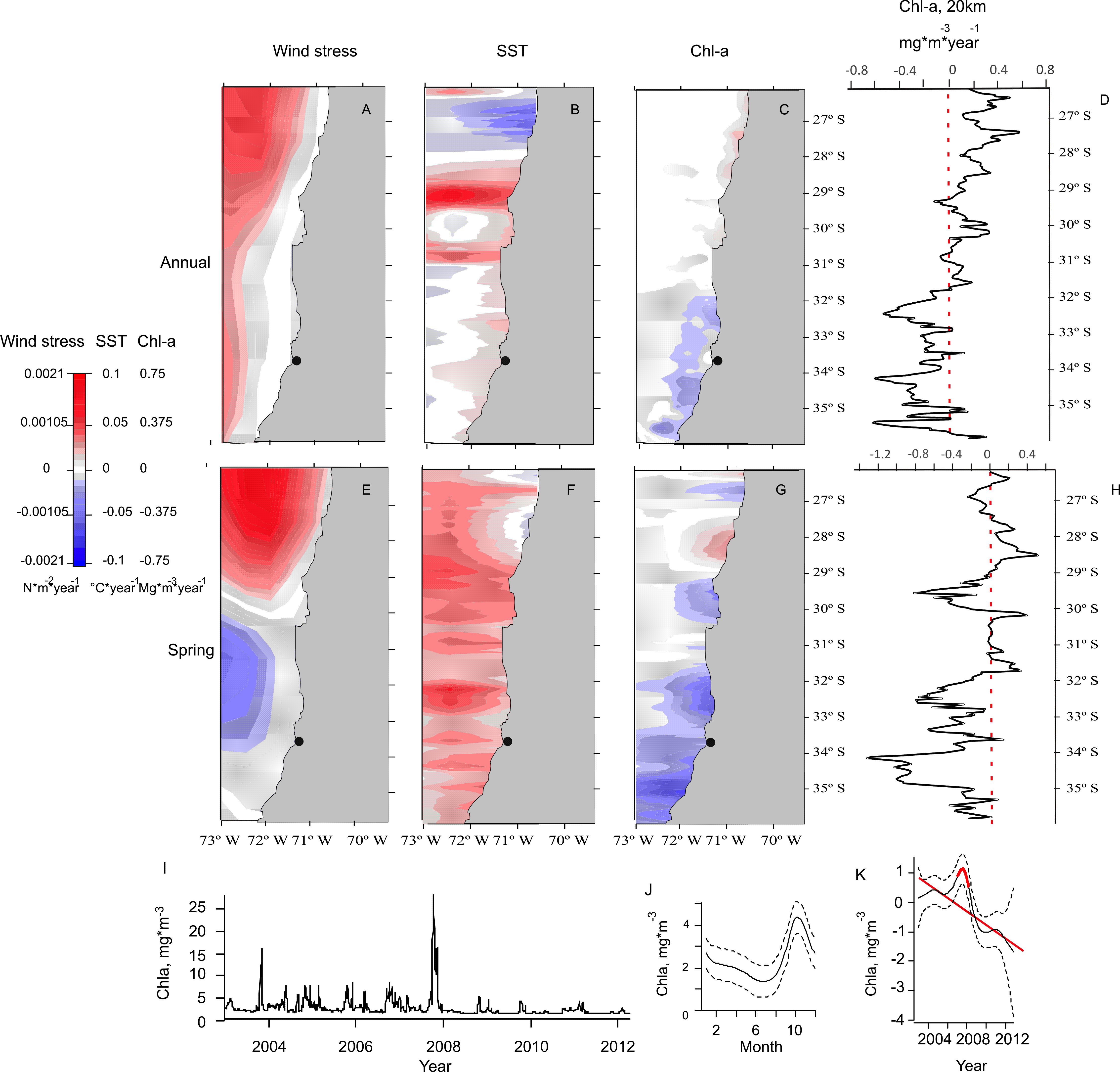}
	\caption{Spatial patterns of long-term trends in meridional wind stress (A, E), SST (B, F) and Chl-a (C, G) for the whole year (A-C) and spring (E-G). Latitudinal trends of Chl-a for the coastal megapixels (0–20 km offshore) for the whole year (D) and spring (H) are also shown, with red dashed lines showing null slopes. Time series of in situ Chl-a at ECIM (black dot on the maps) are represented (I) together with its seasonal (J) and long-term components (K) inferred from the GAMM.}
	\label{fig:fig3}
\end{figure}

The response of SST yearly averages over the study period has been a
coastal warming of around 0.05 °C year\textsuperscript{-1} in the region
south of ca. 29° S and down to 35° S, reaching maximum values of 0.1 °C
year\textsuperscript{-1} just north of Punta Lengua de Vaca (e.g. 30° S,
Fig. 3B). Further north, around 27° S, yearly average temperatures have
decreased at rates of 0.075 °C year\textsuperscript{-1}. In austral
spring months, long-term trends indicate a warming rate of 0.05 °C
year\textsuperscript{-1} along most of the domain, except for a coastal
band north of about 28° S, where surface cooling has taken place (Fig.
3F). Similar cooling patterns were observed for austral winter months
further offshore (\textgreater80 km offshore, S1).

Long-term yearly trends in surface Chl-a emphasize the contrast between
a northern domain, between 26° S and 30° S and a southern domain that
extends to 35° S. Equatorward, yearly primary productivity has increased
at rates of 0.4 mg m\textsuperscript{-3} year\textsuperscript{-1}, while
in the poleward domain it has decreased at faster rates (-0.6 mg
m\textsuperscript{-3} year\textsuperscript{-1}, Fig. 2C-D). A transition
region of little or no temporal trends in overall Chl-a persisted around
30º--31° S (Fig. 3D). Spring months showed a broadly similar pattern
marked by mesoscale spatial structure. Sharp long-term decreases in
Chl-a with rates steeper than -1.2 mg m\textsuperscript{-3}
year\textsuperscript{-1} were widespread in the southern domain,
especially between 34º--35° S (Fig. 3G), while the northern domain
showed increases of about 0.4 mg m\textsuperscript{-3}
year\textsuperscript{-1} , for example between 29º-- 28° S, but other
coastal areas show{~ }little or no change (Fig. 3G). Similar contrasting
trends between the two latitudinal domains were observed for the rest of
the seasons (S1). Daily time series of \emph{in situ} Chl-a at ECIM were
highly consistent with the pattern in primary productivity found at the
southern domain from satellite data. Starting in 2008, the
characteristic spring and fall blooms that were apparent in the time
series between1999-- 2008, had all but disappeared (Fig. 3I, see Wieters
et al., 2003 for previous years), generating a marked long-term decrease
\emph{in situ} Chl-a concentration at this coastal site. Although a
negative linear function significantly fits the long term GAMM trend,
with a slope of -0.22 mg m\textsuperscript{-3} year\textsuperscript{-1}
(p-value \textless0.0001), the temporal trend at the site was more
complex, with a clear shift to overall lower values after a sharp peak
in austral spring 2007-- 2008 (Figs. 3I,K). The seasonal component of
the GAMM was also significant (p-value \textless0.0001) and showed a
marked peak in October (spring) followed by a progressive decrease
through summer and autumn months toward minimum winter values in July (Fig. 3J). 

Differentials in wind stress forcing, SST, and the consequent changes in
Chl-a, between the period before and after 2007, which was the year that
SPA poleward migration apparently accelerated (Schneider et al., 2017),
provide only weak support for the coastal upwelling driven productivity
model. The northern coastal region presented a marked increase of 0.0038
N m\textsuperscript{-2} in wind stress between these two periods, but a
nil or even a slight rise of SST of about 0.005 °C, yet a significant
increase in surface Chl-a of 0.6 mg m\textsuperscript{-3} did occur
(Fig. 4E). At the southern coastal domain, southerly wind stress dropped
0.0002 N m\textsuperscript{-2} between these periods, SST rose 0.25 °C
and Chl-a decreased 1.5 mg m\textsuperscript{-3}, thus closely following
the responses expected from the upwelling-driven productivity model
(Fig. 4F). Further south, at 36.5° S where a long-term monitoring site
is located (Schneider et al., 2017), we did find an increase in
meridional wind stress, as shown in previous studies, although the
magnitude was smaller than 1e-4 N m\textsuperscript{-2}. Surface water
temperature dropped 0.25 °C between these two periods, but instead of
increasing, Chl-a fell 0.75mg m\textsuperscript{-3} (Fig. 3G). 

\begin{figure}
	\centering
	\includegraphics[width=1\linewidth]{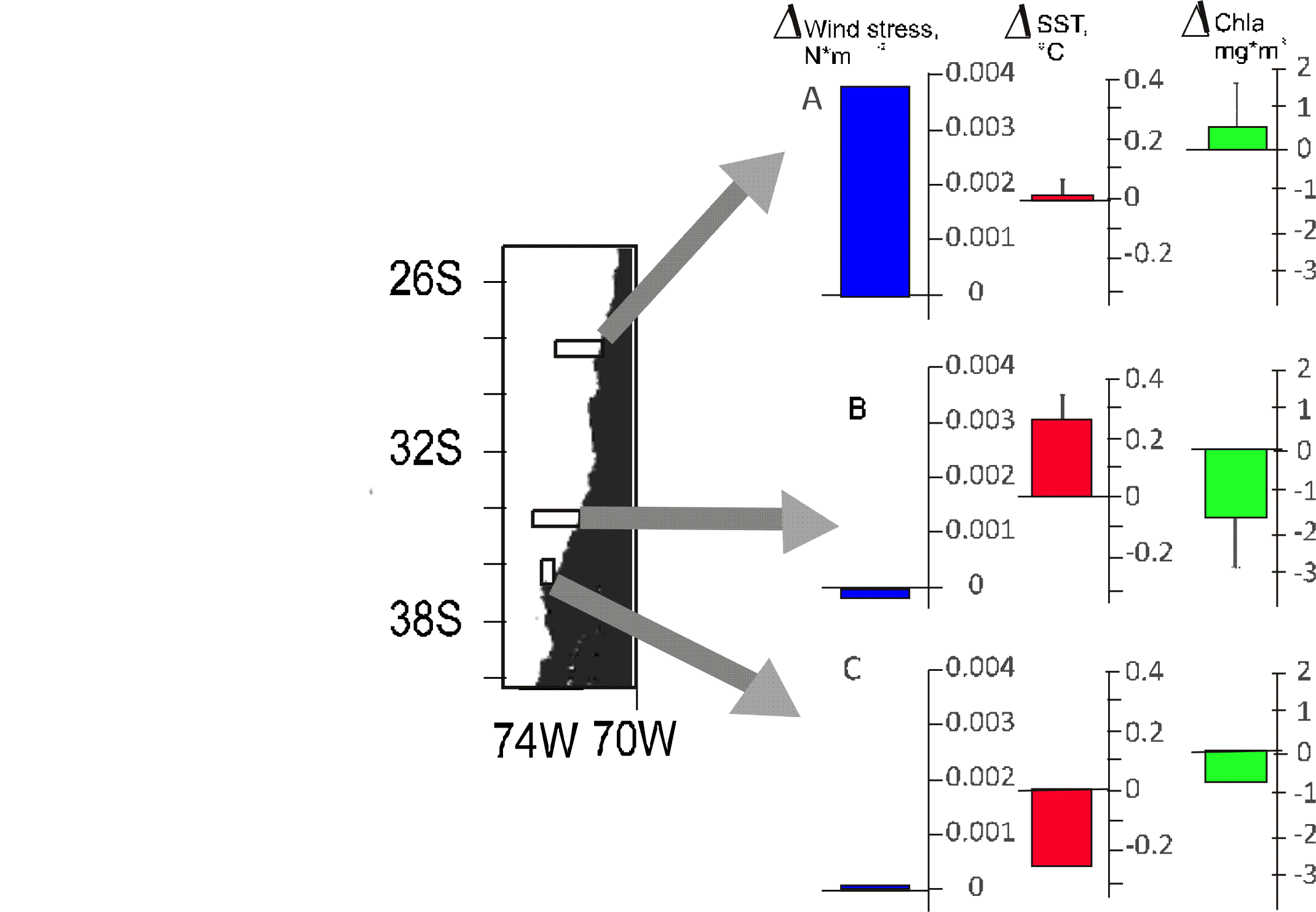}
	\caption{Differences in meridional wind stress, SST and Chl-a between the periods 2003–2007 and 2008–2015 for coastal megapixels at the northern (A) and southern (B) regions marked in the maps by rectangles. In addition, the same temporal differences further south at 36.5° S (station 18) extracted from Schneider et al. (2017) and Aguirre et al. (2018) are shown (C).}
	\label{fig:fig4}
\end{figure}

\section{Discussion}

Here we documented the existence of contrasting long-term trends in wind
forcing and SST along different portions of the Humboldt Upwelling
Ecosystem, and show how these climate-driven changes have translated
into dissimilar domains of surface water productivity in the costal
ocean. At the equatorward domain of the region, north of 30° S,
phytoplankton productivity (Chl-a biomass) has increased in nearshore
habitats since the early 2000's. At the poleward section, between
31º--35° S and over the same time period, phytoplankton productivity has
significantly decreased over an extensive cross-shore region
encompassing the entire continental shelf. The upwelling-driven surface
productivity model provides a partial explanation for the observed
trends in Chl-a productivity in the two domains of the region,
evidencing the complexity of ecosystem responses to altered upwelling
winds (Bakun et al., 2015). The long-term changes in wind forcing are in
remarkable agreement with the temporal and latitudinal changes in two
processes linked to anthropogenic global change in most EBUS, including
HUE. First, we document an increased sea-land temperature gradient (i.e.
Bakun effect) north of 30° S since the early 2000's that has led to
increased upwelling winds in that portion of the region. Second, the
well documented poleward displacement of the SPA beyond 35° S has
increased spring winds south of that latitude but created a gap in
between 31º--35° S where upwelling winds are now significantly weaker.

The regular seasonal changes in the position of the SPA are known to
drive upwelling variability along most of the HUE (Strub et al., 1998;
Ancapichun and Garcés-Vargas 2015). During the second half of the XX
century, the core of the SPA in spring months was located at 33° S
migrating farther south in summer and back up north to its autumn/winter
position around 26° S (Fig. 1A). Yet over the past two decades, and in
agreement with patterns observed for the high pressure systems of the
North and South Atlantic (Kim et al., 2015; He et al., 2017; Reboita et
al., 2019), the spring and, to a lesser extent, summer positions of SPA
have been displaced poleward, reaching{~ }around 35° S and 37° S in
spring and summer, respectively (Figs. 1A). Previous studies using
combinations of ERA Interim reanalysis and GCM multimodel simulations
(Aguirre et al., 2018), together with primary data from QuikScat and
ASCAT scatterometers (Schneider et al., 2017), have also documented the
increased SPA poleward spring displacement, lending support to our
results. The displacement has led to the intensification of spring winds
at the southern-end of HUE (Jacob et al., 2018; Narváez et al., 2019),
and a region of weaker spring winds in central Chile, between 30º--35°S.
Decreased upwelling winds in this extensive region of the Chilean coast
has caused a weak but significant trend to increasing SST, which is
particularly apparent in front of the upwelling hot-spots of Matanzas
(34.1° S), Punta Toro (33.8° S) and at the well-documented upwelling
centre of Punta Lengua de Vaca (30.2° S). As expected from the
upwelling-driven coastal productivity model (e.g. Botsford et al.,
2003), weakening upwelling winds have led to a regional domain with
decreased primary productivity for coastal waters over the past two
decades, which is evidenced by long-term patterns of spring and mean
annual Chl-a from satellite and in situ records. The region of decreased
productivity extends well beyond the coastal zone, especially in spring
months when surface waters with significantly reduced concentration of
Chl-a are found as far as 100 km offshore (Fig. 2B,F). Such an extensive
region of lower annual and spring Chl-a is bound to have effects on
higher trophic levels, as the reduced primary productivity cascades up
the pelagic and benthic food webs (Sydeman et al., 2006; Powell and Xu
2011). Regional displacements of coastal fisheries, top predators and
sea birds to areas of higher productivity should already be taking place
or should be observed in the near future. However, the decreasing Chl-a
rates are not spatially homogeneous within the study region, exhibiting
important and ecologically-significant meso-scale variation (Navarrete
et al., 2005; Valdivia et al., 2015), with small nearshore sectors that
have experienced almost no change (rates of less than --0.1 mg Chl-a /
yr) to regions of very rapid decrease (--1 mg Chla / yr). The long-term
decreasing trends in Chl-a detected from satellite imagery few
kilometres offshore (first pixel 20 km from shore) have also been
observed in time series of in situ chlorophyll-a recorded in water
samples collected from the intertidal zone at ECIM (Fig. 3I-K). Thus,
the domain of decreased productivity includes nearshore waters and it
should therefore affect the rich and heavily exploited intertidal and
subtidal benthic ecosystems (Kefi et al., 2015, Perez-Matus et al., 2017).

Equatorward of 30° S, we show that upwelling-favourable winds have
actually shown significant increases, especially in spring months. Such
wind intensification is consistent with the increased sea-land thermal
differences observed from north of 30° S (Fig. 1), thus we interpret the
spatial pattern as a geographically constrained Bakun effect (Bakun,
1990). Note that upwelling strengthening in the northern domain cannot
be explained by SPA dynamics, as in spring the SPA is located further
south. This spatial decoupling between migration of pressure systems and
Bakun effect has been observed in Monterey Bay, in the California
Current System, where sea-land thermal diurnal breezes cause pronounced
upwelling events regardless of the prevalent forcing exerted by the
North Pacific High off California (Woodson et al., 2007). Increased
winds in our northern region have not led to generally decreasing SST
trends (Fig 2B,F). Long term cooling in the northern region has
apparently occurred only locally (between 27º--28° S), while many
sectors have experienced either no change or even weak warming, probably
due to the complex interaction between the upwelling dynamics, increased
solar radiation (Echevin et al., 2012; Aguirre et al., 2018) and changes
in other processes that modulate thermal conditions and which also
change around this latitude (Tapia et al., 2014). For instance, stronger
stratification by atmospheric warming may lead to upward transport of
warm and already nutrient depleted waters to the surface unable to
trigger conspicuous phytoplankton blooms. This scenario has been
forecasted for the 21\textsuperscript{st} century within the HUE and off
the Northwestern Iberian Peninsula, as increased stratification will
lead to a shallower thermocline (Oyarzun and Brierley, 2018; Sousa et
al., 2020). Even if meridional wind stress increases, a shallow
thermocline may trap wind induced turbulence and currents, thus upwelled
waters may ascend from shallower depths (Pringle 2007). It is worth
noting that our results contrast with earlier analyses of long-term
satellite trends in coastal SST over our study region, which showed a
largely homogenous decreasing trend over the 1982-- 2010 period (Lima
and Wethey, 2012). The difference can be reconciled with the shift in the SPA position observed from 2010 onwards.

Overall, increases or decreases in upwelling winds and the position of
the high-pressure systems do not only alter offshore advection, but also
many other processes that can affect productivity of the coastal ocean
(Bakun et al., 2015). For instance, increased upwelling winds at 36.5° S
(Station 18) due to SPA displacement, have produced sharp cooling of the
water column, but Chl-a productivity has decreased (Fig. 4F, Schneider
et al. 2017; Aguirre et al., 2018; Jacob et al., 2018). It has been
hypothesized that lower precipitations due to droughts linked with the
SPA intensification at that latitude, may be negatively affecting
riverine outflow and reducing primary productivity (Jacob et al., 2018).
Thus, the climatic drivers of upwelling intensity do not only alter
offshore advective transport, but several other factors that also
modulate ocean productivity, and riverine input is one of important one in several regions of EBUS (Thomas and Strubb 2001; Silva et al., 2009; Masotti et al., 2018). 

The driving mechanisms underlying the two contrasting domains in primary
productivity of the coastal ocean along the HUE, as we propose here, can
be associated to anthropogenic global warming. First, although the
dynamics of the SPA are partially driven by the natural variability of
the Pacific Decadal Oscillation (Ancapichun and Garcés Vargas, 2015),
anthropogenic warming appears to be the main cause of the poleward
displacement of high EBUS pressure systems (Lee 1999; Garreaud and
Falvey, 2009; Wang and Cai 2013; Fan et al., 2014; Aguirre et al.,
2018). Second, Bakun effect is a response to the increased sea-land
pressure differential produced by global warming, which should be
intensified at lower latitudes (Bakun, 1990). We show that the Bakun
effect does not change gradually with latitude but exhibits a sharp
discontinuity at about 30° S. The sharp increase of the Bakun effect
equatorward of 30° S is likely related to stronger land warming over the
coastal section of the arid Andean slopes (Vargas et al., 2007). Such
land warming may be weaker to the south of 30° S, where coastal lowlands lay in between the Pacific shores and the Andes. 

Further south and using data from station 18 at 36.5° S, complemented
with wind data from satellite scatterometers, Schneider et al. (2017)
showed an abrupt acceleration of SPA poleward displacement since 2007.
Our analyses showed important differences in wind stress, SST and
especially Chl-a between the beginning of the 2000's and the last decade
of the time series at the 2 domains studied here and the southern end
(station 18) analysed in previous studies (Fig 4 F, G, H). But SPA
latitudinal position changed progressively over time (Fig. 1A) and most
of the time trends detected by the GAMMs all over the region (Fig. 3)
were linear with no evidence of discontinuity. However, a clear
exception was the longterm trend of \emph{in situ} Chl-a at ECIM, which
showed a marked decrease only after a conspicuous Chl-a peak in spring
2007 (Fig. 3 I, K). This temporal pattern at ECIM was mirrored by the
outflow of the Biobío river further south that also dropped after a
large peak in 2007 (Jacob et al., 2018), suggesting large
climaticchanges took place on that year. Cold SST anomalies nearshore
during 2007 up to 29° S were thought to be driven by a change from El
Niño to La Niña conditions or by a northward intrusion of sub-Antarctic
waters (Aravena et al., 2014; Schneider et al., 2017). How these
processes may interact with SPA dynamics to produce more or less
pronounced shifts in oceanographic conditions all along the region remains unclear.

\subsection{Conclusions }

In summary, two main effects of anthropogenic global warming on
upwelling systems seem to be responsible for two contrasting domains of
long-term dynamics in primary productivity of the coast, with limits
between these domains around 30º S to 31° S. To the north, Bakun's
increased sea-land thermal contrast would enhance upwelling-favourable
winds and coastal primary productivity. To the south of the latitudinal
break, there is an ample region of decreasing upwelling winds that
extends down to 34º--35° S due to the poleward displacement of SPA,
where Chl-a concentrations have been drastically reduced. Thus,
anthropogenic climatic trends may generate much more marked and
conspicuous spatial heterogeneity in upwelling intensity and coastal
water productivity than previously thought, which will affect entire
coastal ecosystems by propagation through the food web. Our results
highlight the urgent need for monitoring the coastal ocean to assess
existing trends and forecast future changes in productivity within EBUS,
and in this manner increase our capacity to adapt human activities to the cascading and far reaching effects of global warming.

\section*{  Acknowledgements}

We thank research assistants and students, especially Ivan Albornoz, for
helping in the collection of water samples at ECIM and Mirtala Parragué for managing the \emph{in-situ} Chlorophyll-a dataset. The National Fund for Scientific and Technological Development, FONDECYT (Chile), supported with post-doctoral grants to NW {[}3150072{]}, AO {[}3150425{]} and JB {[}3160294{]}.{ Conversations with Catalina Aguirre and colleagues at ECIM are much appreciated. Additionally, AO was  supported by H2020 Marie Skłodowska-Curie Actions MSCA-IF-2016 {[}746361{]}. SAN was supported by }FONDECYT {[}1160289{]}, Laboratorio Internacional en Cambio Global, LINCGlobal, and CONICYT PIA/BASAL FB0002.

\section*{  References}

Aguirre, C., Pizarro, O., Strub, P.T., Garreaud, R. Barth, J.A. 2012.
Seasonal dynamics of the near-surface alongshore flow off central Chile. J. Geophys. Res., 117, C01006, doi:10.1029/2011JC007379.

Aguirre, C., Garcia-Loyola, S., Testa, G., Silva, D. Farias, L. 2018.
Insight into anthropogenic forcing on coastal upwelling off
south-central Chile. Elementa-Sci. Anthro., 6, 59. doi:10.1525/elementa.314

Aiken, C.M., Navarrete, S.A., Pelegrí, J.L. 2011. Potential changes in
larval dispersal and alongshore connectivity on the central Chilean
coast due to an altered wind climate. J. Geophys. Res., 116, G04026,
doi:10.1029/2011JG001731.

Anabalón, V., Morales, C.E., González, H.E., Menschel, E., Schneider,
W., Hormazabal, S., Valencia, L., Escribano, R. 2016.
Micro-phytoplankton community structure in the coastal upwelling zone
off Concepción central Chile: Annual and inter-annual fluctuations in a highly dynamic environment. Prog. Oceanogr., 149, 174-188.

Ancapichun, S., Garcés- Vargas, J. 2015. Variability of the Southeast
Pacific Subtropical Anticyclone and its impact on sea surface temperature off north-central Chile. Ciencias Marinas, 411, 1--20.

Aravena, G., Broitman, B., Stenseth, N.C. 2014. Twelve years of change
in coastal upwelling along the Central-Northern coast of Chile:
spatially heterogeneous responses to climatic variability. PLoS ONE 92, e90276. doi:10.1371/journal.pone.0090276.

Bakun, A. 1990. Global climate change and intensification of coastal ocean upwelling. Science, 247, 198--201.

Bakun, A., David, B., Field, W., Redondo-Rodriguez, A. Weeks, S.J. 2010.
Greenhouse gas, upwelling-favorable winds, and the future of coastal
ocean upwelling ecosystems. Global Change Biology 16, 1213--1228, doi: 10.1111/j.1365-2486.2009.02094.x.

Bakun, A., Black, B.A., Bograd, S.J., Garcia-Reyes, M., Miller, A.J.,
Rykaczewski, R.R., Sydeman, W.J. 2015. Anticipated effects of climate
change on coastal upwelling ecosystems. Current Climate Change Reports, 1, 85-93.

Blanco, J. L., Carr, M. E., Thomas, A. C., Strub, P. T. 2002.
Hydrographic conditions off northern Chile during the 1996 -- 1998 La
Niña and El Niño events. J. Geophys. Res., 107C3, doi:10.1029/2001JC001002.

Botsford, L.W., Lawrence, C.A., Dever, E.P., Hastings, A., Largier, J.
2003. Wind strength and biological productivity in upwelling systems: an idealized study. Fisheries and Oceanography 12:4/5, 245--259.

Chavez, F.P., Messie, M. 2009. A comparison of Eastern Boundary Upwelling Ecosystems. Prog. Oceanogr. 83:80-96.

Chen, C. 2000. Generalized additive mixed models, Communications in
Statistics - Theory and Methods, 29:5-6, 1257-1271, DOI: 10.1080/03610920008832543

Chenillat, F., Riviere, P., Capet, X., Franks, P.J.S., Blanke, B. 2013.
California coastal upwelling onset variability: cross-shore and
bottom-up propagation in the planktonic ecosystem. PLoS One 85, e62281

Cury, P., Roy, C., Faure, V. 1998. Environmental constraints and pelagic
fisheries in upwelling areas: the Peruvian puzzle. Afr. J. Mar. Sci., 19,159-167.

Cushing, D.H. 1971. Upwelling and the production of fish. Advan. Mar. Biol., 9, 255-334.

Dee, D.P., Uppala, S.M., Simmons, A.J., Berrisford, P., Poli, P.,
Kobayashi, S., Andrae, U., Balmaseda, M.A., Balsamo, G., Bauer, P.,
Bechtold, P., Beljaars, A.C.M., van de Berg, L., Bidlot, J., Bormann,
N., Delsol, C., Dragani, R., Fuentes, M., Geer, A.J., Haimberger, L.,
Healy, S.B., Hersbach, H., Holm, E.V., Isaksen, L., Kallberg, P.,
Kohler, M., Matricardi, M., McNally, A.P., Monge-Sanz, B.M., Morcrette,
J.J., Park, B.K., Peubey, C., de Rosnay, P., Tavolato, C., Thepaut,
J.N., Vitart, F. 2011. The ERA-Interim reanalysis: configuration and
performance of the data assimilation system. Q. J. Royal{~ }Meteorol. Soc., 137, 553-597.DOI:10.1002/qj.828

Demarcq, H. 2009. Trends in primary production, sea surface temperature
and wind in upwelling systems 1998-2007. Prog. Oceanogr., 83, 376--385. doi: 10.1016/j.pocean.2009.07.022

Echevin, V., Goubanova, K., Belmadani, A., Dewitte, B. 2012. Sensitivity
of the Humboldt system to global warming: a downscaling experiment of
the IPSL-CM4 model. Climate Dyn., 38, 761-774. doi:10.1007/s00382-011-1085-2.

Fan, Y., Lin, S., Griffies, S. M., Hemer, M. A. 2014. Simulated global
swell and wind-sea climate and their responses to anthropogenic climate
change at the end of the twenty-first century. J. Climate, 27, 3516--3536

García-Reyes, M., Largier, J. 2010. Observations of increased
wind-driven coastal upwelling off central California. J. Geophys. Res., 115, C04011. doi: 10.1029/2009jc005576

Garreaud, R.D., Falvey, M. 2009. The coastal winds off western
subtropical South America in future climate scenarios. Int. J. Climatol., 29, 543--554.

Gómez-Letona, M., Ramos, A.G., Coca, J. Arístegui, J. 2017. Trends in
primary production in the Canary Current Upwelling System---a regional
perspective comparing remote sensing models. Front. Mar. Sci., 4, 370.doi: 10.3389/fmars.2017.00370

Harding, L.W., Gallegos, L.C., Perry, E.S., Miller, W.D., Adolf, J.E.,
Mallonee, M.E., Paerl, H.W. 2016. Long term trends in nutrient and phytoplankton in Chesapeake Bay. Estuar. Coast., 39, 664--681.

Haye, P.A., Segovia, N.I., Muñoz-Herrera, N.C., Gálvez, F.E., Martínez,
A., Meynard, A., Pardo-Gandarillas, M.C., Poulin, E., Faugeron, S. 2014.
Phylogeographic structure in benthic marine invertebrates of the
Southeast Pacific coast of Chile with differing dispersal potential. PLoS ONE 92, e88613. doi:10.1371/journal.pone.0088613

He, C., Wu, B., Zou, L., Zhou, T. 2017. Responses of the summertime
subtropical anticyclones to global warming. J. Climate, 30, 6465-6479.

Hormazabal, S., Shaffer, G., Letelier, J., Ulloa, O. 2001. Local and
remote forcing of sea surface temperature in the coastal upwelling
system off Chile. J. Geophys. Res., Oceans, 106, 16657--16671. doi:10.1029/2001JC900008.

Jacob, B.J., Tapia, F.J., Quiñones, R.A., Montes, R., Sobarzo, M.,
Schneider, W.\ldots{} Gonzalez, H.E. 2018. Major changes in diatom
abundance, productivity, and net community metabolism in a windier and
dryer coastal climate in the southern Humboldt Current. Prog. Oceanogr., 168, 196-209.

Kefi, S., Berlow, E.L., Wieters, E.A., Joppa, L.N., Wood, S.A., Brose,
U., Navarrete, S.A. 2015. Network structure beyond food webs: mapping
non-trophic and trophic interactions on Chilean rocky shores. Ecology,
961, 291-303.

Kim, Y.H., Kim, M.K., Lau, W.K.M., Kim, K.M., Cho, C.H. 2015. Possible
mechanism of abrupt jump in winter surface air temperature in the late
1980s over the Northern Hemisphere, J. Geophys. Res., Atmosphere, 120,
12474--12485, doi:10.1002/2015JD023864.

Lachkar, Z., Gruber, N. 2012. A comparative study of biological
production in eastern boundary upwelling systems using an artificial
neural network. Biogeosciences 9, 293--308.

Lee, S. 1999. Why are the climatological zonal winds easterly in the
Equatorial upper troposphere? J. Atmos. Sci., 56,1353-1363.

Lima, F. P., Wethey, D. S., 2012. Three decades of high - resolution sea
coastal sea surface temperature reveal more than warming. Nat. Commun.,
3, 704.

Mann, K. H., Lazier, J. R. N. 2006. Vertical structure in coastal
waters: coastal upwelling regions. Pages 139-179 in Mann, K. H., Lazier,
J. R. N. Dynamics of Marine Ecosystems. Blackwell Science, Cambridge.

Marshall, G. J. 2003. Trends in the Southern Annular Mode from
observations and reanalyses. J. Clim. 16: 4134-4143.

Masotti, I., Aparicio-Rizzo, P., Yevenes, M.A., Garreaud, R., Belmar,
L., Farías, L. 2018. The influence of river discharge on nutrient export
and phytoplankton biomass off the Central Chile Coast 33°--37°S:
seasonal cycle and interannual variability. Front. Mar. Sci., 5(423).
doi: 10.3389/fmars.2018.00423.

McGregor, H. V., Dima, M., Fischer, H. W., Mulitza, S. 2007. Rapid 20th
century increase in coastal upwelling off Northwest Africa. Science 315:
637-639.

Minetti, J.L., Vargas, W.M., Poblete, A.G., Mendoza, E.A. 2009.
Latitudinal positioning of the subtropical anticyclone along the Chilean
coast. Aust. Meteorol. Ocean., 58, 107-117.

Narváez, D.A., Vargas, C.A., Cuevas, L.A., García-Loyola, S.A., Lara,
C., Segura, C., Tapia, F.J., Broitman, B.R. 2019. Dominant scales of
subtidal variability in coastal hydrography of the Northern Chilean
Patagonia. J. Mar. Syst., 193, 59-73.

Navarrete, S.A., Wieters, E.A., Broitman, B.R., Castilla, J.C. 2005.
Scales of benthic-- pelagic coupling and the intensity of species
interactions: from recruitment limitation to top-down control. Proc.
Natl. Acad. Sci. U. S. A., 102, 18046-18051.

Nguyen, H., Evans, A., Lucas, C., Smith, I., Timbal, B. 2013. The Hadley
Circulation in reanalyses: climatology, variability, and change. J.
Climate, 3357-3375.

Oyarzun, D., Brierley, C.M. 2018. The future of coastal upwelling in the
Humboldt current from model projections. Climate Dyn., 52, 599-615.

Pardo, P.C., Padín, X.A., Gilcoto, M., Farina-Busto, L., Perez, F.F.
2011. Evolution of upwelling systems coupled to the long-term
variability in sea surface temperature and Ekman transport. Climate Res.
48, 231-246.

Pauli, D., Christensen, V. 1995. Primary production required to sustain
global fisheries. Nature, 374, 255-257

Perez-Matus, A., Carrasco, S. A., Gelcich, S., Fernandez, M. Wieters,
E.A. 2017. Exploring the effects of fishing pressure and upwelling
intensity over subtidal kelp forest communities in Central Chile.
Ecosphere 85, e01808. 10.1002/ecs2.1808.

Powell Jr., A.M., Xu, J. 2011. Abrupt climate regime shifts, their
potential forcing and fisheries impacts. Atmos. Climate Sci., 1,
33-47.

Previdi, M., Liepert, B.G. 2007. Annular modes and Hadley cell expansion
under global warming. Geophys. Res. Lett., 34, L22701,
doi:10.1029/2007GL031243.

Pringle, J.M. 2007. Turbulence avoidance and the wind-driven transport
of plankton in the surface Ekman layer. Cont. Shelf Res., 27,
670-678.

R Core Team 2019. R: A language and environment for statistical
computing. R Foundation for Statistical Computing, Vienna, Austria. URL
https://www.R-project.org/.

Reboita, M. S., Ambrizzi, T., Silva, B.A., Pinheiro, R.F., da Rocha,
R.P. 2019. The South Atlantic Subtropical Anticyclone: present and
future climate. Front. Earth Sci., 7, 8. doi: 10.3389/feart.2019.00008.

Salas, S., Chuenpagdee, R., Charles, Seijo, A.J.C. editors. 2011.
Coastal fisheries of Latin America and the Caribbean. FAO, Rome.

Schneider, W., Donoso, D., Garces-Vargas, J., Escribano, R. 2017.
Water-column cooling and sea surface salinity increase in the upwelling
region off central-south Chile driven by a poleward displacement of the
South Pacific High. Prog. Oceanogr., 151, 38-48. Silva, A., Palma, S.,
Oliveira, P.B., Moita, M. T. 2009. Composition and interannual

variability of phytoplankton in a coastal upwelling region Lisbon Bay,
Portugal. J. Sea Res., 62, 238--249.

Simpson, G.L. 2018 Modelling palaeoecological time series using
generalised additive models. Front. Ecol. Evol., 6, 149, doi:
10.3389/fevo.2018.00149.

Smayda, T.J. 2000 Ecological features of harmful algal blooms in coastal
upwelling ecosystems. Afr. J. Mar. Sci., 22, 219--253.

Sousa, M. C., Ribeiro, A., Des, M., Gomez-Gesteira, M., deCastro, M.,
Dias, J. M. 2020. NW Iberian Peninsula coastal upwelling future
weakening: Competition between wind intensification and surface heating.
Sci. Total Environ 703 (134808)

Strub, P.T., Mesias, J., Montecino, B.V., Rutllant, J.A., Marchant, S.S.
1998. Coastal ocean circulation off western South America, in: The Sea,
Vol. 11, edited by A.R. Robinson and K.H. Brink, pp. 273-314, John
Wiley, New York, 1998.

Sydeman, W.J., Bradley, R.W., Warzybok, P., Abraham, C.L., Jahncke, J.,
Hyrenbach, K. D., Kousky, V., Hipfner, J.M., Ohman, M. D. 2006.
Planktivorous auklet Ptychoramphus aleuticus responses to ocean climate,
2005: Unusual atmospheric blocking?. Geo. Res. Lett., 33, L22S09,
doi:10.1029/2006GL026736

Sydeman, W.J., García-Reyes, M., Schoeman, D.S., Rykaczewski,
R.R.,Thompson, S.A., Black, B.A., Bograd, S.J. 2014. Climate change and
wind intensification in coastal upwelling ecosystems. Science, 345,
77--80. doi: 10.1126/science.1251635.

Tapia, F.J., Largier, J.L., Castillo, M., Wieters, E.A., Navarrete, S.A.
2014. Latitudinal discontinuity in thermal conditions along the
nearshore of Central-Northern Chile. PLoS ONE, 910, e110841.
doi:10.1371/journal.pone.0110841

Thomas, A., Strub, P.T. 2001. Cross-shelf phytoplankton pigment
variability in the California Current. Cont. Shelf Res., 21,
1157--1190.

Valdivia, N., Aguilera, M.A., Navarrete, S.A. Broitman, B.R. 2015.
Disentangling the effects of propagule supply on the spatial structure
of a rocky shore metacommunity. Mar. Ecol. Prog. Ser., 538, 67-79.

Varela, R., Álvarez, I., Santos, F., DeCastro, M., Gómez-Gesteira, M.
2015. Has upwelling strengthened along worldwide coasts over 1982-2010?
Sci. Rep., 5, 15. doi: 10.1038/srep10016.

Vargas, G., Pantoja, S., Rutllant, J.A., Lange, C.B. Ortlieb, L. 2007.
Enhancement of coastal upwelling and interdecadal ENSO-like variability
in the Peru-Chile Current since late 19th century. Geophys. Res. Lett.,
34, L13607, doi:10.1029/2006GL028812.

Wang, G., Cai, W. 2013. Climate change impact on the 20th century
relationship between the southern annular mode and global mean
temperature. Sci. Rep., 3, 2039, doi:10.1038/srep02039.

Wang, D., Gouhier, T.C., Menge, B.A., Ganguly, A.R. 2015.
Intensification and spatial homogenization of coastal upwelling under
climate change. Nature, 518, 390--394. doi:10.1038/nature14235.

Weller, R.A. 2015. Variability and trends in surface meteorology and
air-sea fluxes at a site off Northern Chile. J. Climate, 28,
doi:10.1175/JCLI-D-14-00591.1.

Wieters, E.A., Kaplan, D.M., Navarrete, S.A., Sotomayor, A., Largier,
J.L., Nielsen, K.J.,

Veliz, F. 2003. Alongshore and temporal variability in chlorophyll a
concentration in Chilean nearshore waters. Mar. Ecol. Prog. Ser., 249,
93--105. doi:10.3354/meps249093.

Wieters, E.A., Salles, E., Januario, S.M., Navarrete, S.A. 2009. Refuge
utilization and preferences between competing intertidal crab species.
J. Exp. Mar. Biol. Ecol., 374, 37-44.

Wood S.N. 2006. Generalized Additive Models: An Introduction with R.
Chapman and Hall/CRC Press.

Woodson, C.B., Eerkes-Medrano, D.I., Flores-Morales, A., Foley, M.M.,
Henkel, S.K., Hessing-Lewis, M., Jacinto, D., Needles, L., Nishizaki,
M.T., O'Leary, J., Ostrander, C.E., Pespeni, M., Schwager, K.B.,
Tyburczy, J.A., Weersing, K.A., Kirincich, A.R., Barth, J.A., McManus,
M.A., Washburn, L. 2007. Local diurnal upwelling driven by sea breezes
in northern Monterrey Bay. Cont. Shelf Res., 27, 2289--2302.

Yentsch, C.S., Menzel, D.W. 1963. A method for the determination of
phytoplankton chlorophyll and phaeophytin by fluorescence. Deep-Sea
Res., 10, 221-231.

\pagebreak
\appendix
\label{sec:Appendix}
\section*{Appendix}
\includepdf[pages=1]{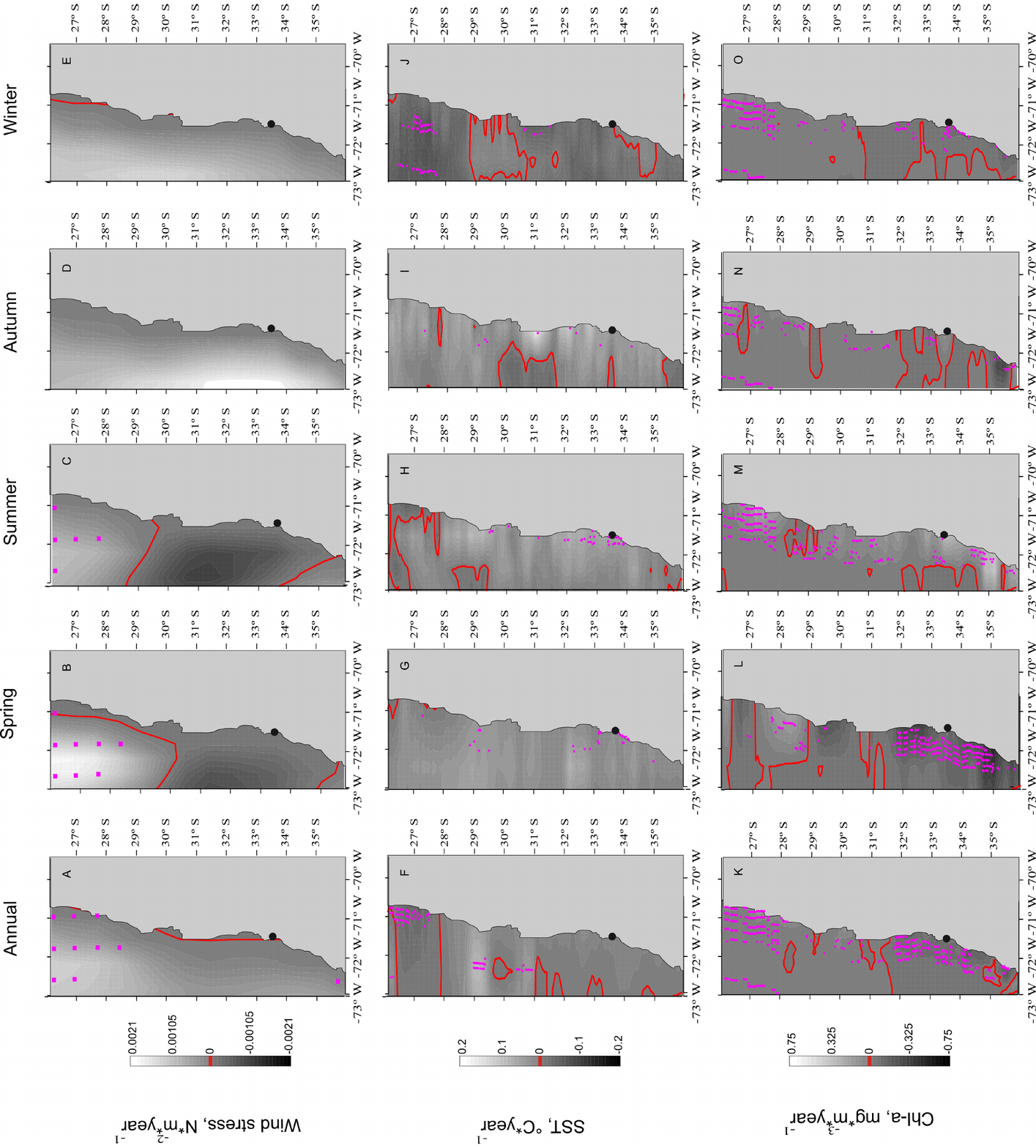}

\end{document}